\documentstyle[12pt]{article}
\newlength{\eff}

\def\d{\partial}
\def\q3{QED$_{2+1}$}
\def\frac#1#2{{{\displaystyle #1} \over {\displaystyle #2}}}

\newcommand{\bc}{\begin{center}}
\newcommand{\ec}{\end{center}}

\begin{document}

\large
\thispagestyle{empty}
\begin{flushright}                      FIAN/TD/94-09 \\
                                        hep-th/9410064\\
                                        August 1994

\vspace{3ex}
\end{flushright}
\bc
\normalsize

\vspace{2cm}
{\LARGE\bf QED$_{2+1}$ with Nonzero Fermion Density

and Quantum Hall Effect}

\vspace{1cm}

{\Large Vadim  Zeitlin$^\dagger$}

\vspace{0.5cm}
{\large Department of Theoretical Physics,

\medskip
 P.~N.~Lebedev Physical Institute,

\medskip
Leninsky prospect 53, 117924 Moscow, Russia}

\vspace{1.5cm}
\ec

\centerline{{\Large\bf Abstract}}
\xiipt

\begin{quote}
A general expression for the conductivity in the \q3 with nonzero fermion
density in the uniform magnetic field is derived. It is shown that the
conductivity is entirely determined by the Chern-Simons coefficient:
$\sigma_{ij}=\varepsilon_{ij}~{\cal C}$ and is a step-function of the
chemical
potential and the magnetic field.

\end{quote}

\vfill
$^\dagger$ E-mail address: zeitlin@lpi.ac.ru
\newpage
\normalsize
\setcounter{page}{1}

We shall present here a simple relativistic model
possessing a quantum Hall-like conductivity
-- (Maxwell) quantum electrodynamics on the plane with
nonzero fermion density.
Using the general properties of the \q3 an expression for the conductivity
will be derived. In this model a transverse conductivity arises owing to
the induced  Chern-Simons term,
which is generated dynamically in the one-loop
polarization operator.
The quantization of the Chern-Simons coefficient at a
nonzero chemical potential and magnetic field manifests itself in the
quantization of
the conductivity.
The latter is a step-function of the number of the filled
Landau levels.

\bigskip
We shall consider (2+1)-dimensional QED with a nonzero chemical
potential.  Its Lagrangian reads

\bigskip
        $$
         L = -\frac14 F_{\mu\nu}F^{\mu\nu} +\bar{\psi}(\imath {\partial
        \kern-0.5em/} + e {A\kern-0.5em/}  + \gamma_0 \mu -m)\psi \eqno{(1)}
         $$

\bigskip
\noindent
with standard notations \cite{djt}:
$F_{\mu\nu}=\partial_\mu A_\nu - \partial_\nu A_\mu$,
the magnetic field $B$ is defined as $B = \partial_1A_2 - \partial_2A_1$,
$\gamma$-matrices are Pauli matrices, $\gamma_0 = \sigma_3$,
$\gamma^{1,2}=\imath\/ \sigma_{1,2}$, ~$\mu$ is the chemical
potential (by introducing the chemical potential in this way one has
to modify the
$\imath\epsilon$--prescription in the fermion Green function
\cite{s,ceo}). In this model the fermion mass term violates ${\cal P}$- and
${\cal T}$-parity, thus the Chern-Simons term may be generated
dynamically despite the fact
it is not present in the bare Lagrangian \cite{djt}.

In a strong background magnetic filed the current $\tilde{I}$ induced by the
perturbing electric field may be written as a linear response function:

\bigskip
        $$
        \tilde{I}_\mu(x) = \int d^3\! x' \Pi_{\mu\nu}(x - x'|B,\mu)
        \tilde{A}^\nu(x')  \quad ,
        \eqno{(2)}
        $$

\bigskip
\noindent
$\Pi_{\mu\nu}$ is the polarization operator, $\d_i \tilde{A}^0(x) =
E_i, \quad \tilde{A}_i(x)=0$.

Using the definition of the conductivity $\sigma_{ij} = \left. \frac{\partial
I_i}{\partial E_j} \right|_{E \rightarrow 0} $
one may rewrite Eq.~(2) as follows:

\bigskip
        $$
        \sigma_{ij} =
        \imath \/\/ \left. \frac{\d \Pi_{0i}(p)}{\d p_j} \right|_{p
        \rightarrow 0} \quad.
        \eqno{(3)}
        $$

\bigskip
\noindent
Therefore, the calculation of the conductivity is reduced to the
calculation of  the
polarization operator in the QED$_{2+1}$ with nonzero fermion density with
some uniform magnetic field.  Before the calculation of
the polarization operator
it is worth to obtain its general tensor structure. As it is shown in
Appendix, the polarization operator may be written in the following form:

\bigskip
        $$
        \Pi_{\mu\nu}(p) =
        \left(
        g_{\mu\nu} - \frac{p_\mu p_\nu}{p^2}               \right){\cal A}
        +\left(
        \frac{p_\mu p_\nu}{p^2} -
        \frac{p_\mu u_\nu + u_\mu p_\nu}{(pu)} +
        \frac{u_\mu u_\nu}{(pu)^2} p^2                          \right)
        {\cal B}
        + \imath \varepsilon_{\mu\nu\alpha}p^\alpha
        {\cal C}~.
        \eqno{(4)}
        $$

\bigskip
\noindent
The first tensor structure in Eq.~(4) is the standard vacuum term, the
second one is usually associated to the finite-temperature
effects \cite{frad}
and the last one is the induced Chern-Simons term \cite{djt}.

Since the scalars ${\cal A}$, ${\cal B}$ and ${\cal C}$ are the
functions of
$\mu, ~B, ~p_0$ and ${\bf p}^2$ and $\Pi_{\mu\nu}$ is finite in the
$p \rightarrow 0$
limit, the only term that may survive in the expression for the
conductivity,
Eq.~(3), is the Chern-Simons term

\bigskip
        $$
        \sigma_{ij} =
        \varepsilon_{ij} \sigma =  \varepsilon_{ij} {\cal C}
        \eqno{(3')}
        $$

\bigskip
\noindent
and our task is reduced to the calculation of the Chern-Simons
coefficient in
the limit $p_0=0, ~{\bf p}^2 \rightarrow 0$.  From the other hand,
it is well-known that a nonzero contribution to the Chern-Simons
coefficient
${\cal C}$ in the  $p \rightarrow 0$ limit arises on the one-loop
level only \cite{cs}.

Now the problem is sufficiently simplified: to get the whole answer
for the conductivity we may calculate one-loop Chern-Simons term
only. It is possible to calculate this directly\footnote{In Ref.
\cite{hos} the one-loop polarization operator for the \q3 with the uniform
external field was calculated directly for the particular choises of the
chemical potential corresponding to the vacuum and the filled lowest Landau
level.  The results for the Chern-Simons coefficient for these cases
coincide with those obtained here and in Ref. \cite{z1}.  There is
a discrepancy in the calculation of the coefficient ${\cal B}$ in Ref.
\cite{hos} and Ref.  \cite{z1} which does not affect the results presented
here.} since the one-loop polarization operator may be written as

\bigskip
        $$
        \Pi_{\mu\nu}(p|\mu,B) = \imath  e^2 {\rm tr} \int\!  d^3q \gamma_\mu
        G(p+q|\mu,B) \gamma_\nu G(q|\mu,B)
        \eqno{(5)}
        $$

\bigskip
\noindent
and the corresponding expression for the
fermion Green function $G(p|\mu,B)$ is known \cite{z2}. At the same time,
$\Pi_{\mu\nu}(p)$ is:

\bigskip
        $$
        \Pi_{\mu\nu}(x,x') =
        \imath  ~  \frac{\delta <j_\mu (x)>}{\delta A_\nu (x')}
        \eqno{(6)}
        $$

\bigskip
\noindent
and its components  $\Pi_{0j}$, ~($j=1,2$)   in the static limit are:

\bigskip
        $$
        \Pi_{0j}(p \rightarrow 0) = \imath  ~e \varepsilon_{ij}p_i\frac{\d
        \rho}{\d B} \quad ,
        \eqno{(6')}
        $$

\bigskip
\noindent
$\rho$ is the fermion density, $j_o = e \rho$.

The simplest way to
calculate the fermion density in the \q3 with magnetic
field is based on the index
theorem \cite{n}:  the fermion number $N$ is proportional to a difference
between the number of positive and negative energy levels.
At $T=0,~ \mu \ne
0$ the fermion number $N$ is:

\bigskip
        $$
        N = - {1 \over 2} \sum_k {\rm sign} (\varepsilon_k) +
        \sum_k
        \left(
        \theta(\varepsilon_k)\theta(\mu - \varepsilon_k) -
        \theta(- \varepsilon_k)\theta(\varepsilon_k - \mu)
                        \right)\quad,
        \eqno{(7)}
        $$

\bigskip
\noindent
$\varepsilon_k$ are the energy levels. In the \q3 with a background uniform
magnetic field $B$ the fermion energy spectrum (Landau levels) is discrete,

\bigskip
        $$
        p_0 = - m~{\rm sign} (eB), \quad p_0 = \pm \sqrt{m^2 +2 |eB|n},\quad
        n = 1,2,3, \dots
        \eqno{(8)}
        $$

\bigskip
\noindent
(note the asymmetry of the spectrum). Using Eq.~(7) and taking into account
the degeneracy of the Landau levels $\frac{|eB|}{2\pi}$,  the fermion density
may be
written as follows ($eB > 0$):

\bigskip
        $$
        \rho (B,\mu) = \frac{eB}{4\pi} +
        \left\{
        \begin{array}{cl}
        \frac{eB}{2\pi}
        \left[
        \frac{\mu^2 -m^2}{2eB}          \right] ,\quad     & \mu>\/m;
        \\ {}\\
        0,                   \quad             & |\mu|<m;\\
        {}\\
        {}-
        \frac{eB}{2\pi}
        \left(
        1 + \left[
        \frac{\mu^2 - m^2}{2eB}      \right]    \right) ,\quad &\mu<- m,
        \end{array}\right.
        \eqno{(9)}
        $$

\bigskip
\noindent
where $[ \dots ]$ denotes the integral part.

Now the conductivity may be written as:

\bigskip
        $$
        \sigma = \frac{e^2}{4\pi} + \left\{
        \begin{array}{cl}
        \frac{e^2}{2\pi}
        \left[
        \frac{\mu^2 -m^2}{2eB}        \right],
        \quad &\mu > m;\\
        {}\\
        0,                   \quad             & |\mu|<m;\\
        {}\\
        {}-
        \frac{e^2}{2\pi}
        \left(
        1 + \left[
        \frac{\mu^2 - m^2}{2eB}           \right]    \right) ,
        \quad     & \mu<\/-m.
        \end{array}\right.
        \eqno{(10)}
        $$

\bigskip
It is easy to see that the conductivity $\sigma$ as a function of $\mu$ at a
constant magnetic field or as a function of $B$ at a
fixed chemical potential
is a step-function, therefore we have the integer quantum Hall effect.
At the
same time, for these two cases the jumps of the conductivity are
accompanied by sharp
changes of density, thus we have a "naive" quantum Hall effect.

{}From the other hand we may consider an equation

\bigskip
        $$
        \rho (B,\mu) = \bar{\rho} \quad,
        \eqno{(11)}
        $$

\bigskip
\noindent
$\bar{\rho}$ is constant. This equation has
an infinite set of solutions, namely

\bigskip
        $$
        eB_n = \frac{4\pi\bar{\rho}}{2n+1},
        \quad n=0,1,2, \dots ,
        \eqno{(12)}
        $$

\bigskip
\noindent
see Figure 1. The conductivity of these states (of equal density) is
quantized, $\sigma_n \sim (2n+1)$.

In all above-described schemes  step-like variations of the conductivity
are connected to filling (emptying) of the successive Landau levels: a
discrete fermion spectrum (as well as equal degeneracy of Landau levels)
leads to the quantization of the conductivity in the \q3.

The procedure described may be applied also for the calculation of
the conductivity in the QED$_{3+1}$ in the plane orthogonal to the
direction of
magnetic field. At zero temperature fermion density in QED$_{3+1}$ is
a continuous (but not smooth) function of $B$ and $\mu$ \cite{pz}.
This leads
to the relativistic Schubnikov - de Haas oscillations of conductivity.

In conclusion, we want to stress that approach used in this paper for
the calculation of the conductivity, Eqs.~(3),~(6$'$) is valid for nonzero
electric field, too, but the calculation of the fermion density for
$\mu,~B$ and
$E \ne 0$ is more complicate .

\bigskip
\section*{Acknowledgments}
I am grateful to  A.~Karlhede, G.~Semenoff, and  I.~V.~Tyutin  for
the illuminating discussions and to
 L.~Brink and A.~Niemi
for their kind hospitality at the Institute of Theoretical Physics,
Chalmers University of Technology and the Institute for Theoretical Physics,
Uppsala University, where a part of this work has been done. This  work was
supported in part by the Russian Fund of Fundamental Research Grant $N^o$
67123016 and Grant MQM000 from the International Science Foundation.

\section*{Appendix}

To find a general tensor structure of the polarization operator
$\Pi_{\mu\nu}(p)$  in the \q3 we shall write its eigenvector
decomposition (see, e.g. \cite{shab}):

\bigskip
        $$
        \Pi_{\mu\nu}(p) = \sum_{i=1,2} \kappa_i \frac{b_\mu^{(i)}
        b_\nu^{(i)*}} {b_\alpha^{(i)} b^{\alpha(i)*}} \quad ,
        \eqno{(A1)}
        $$

\bigskip
\noindent
the third eigenvector with zero eigenvalue is $p_\mu$ (due to the gauge
invariance    $\Pi_{\mu\nu}(p)$ is transversal, $\Pi_{\mu\nu}(p)~p^\nu=0$).

The eigenvectors $b^{(i)}$ may be chosen as follows:

\bigskip
        $$
        b^{(i)}_\mu = \varepsilon_{\mu\alpha\beta}u^\alpha p^\beta +
        \imath \alpha_i (p_\mu - u_\mu \frac{p^2}{(pu)} ),
        \quad  b^{(i)}_\mu p^\mu \equiv 0     \quad ,
        \eqno{(A2)}
        $$

\bigskip
\noindent
$u^\mu$ is the $3$-velocity of the medium \cite{frad}, $u^\mu = (1,0,0)$.

The orthogonality condition   $b_\alpha^{(1)} b^{\alpha(2)*}=0$ fixes one of
the coefficients  $\alpha_i$, ~~$\alpha_1\alpha_2^* = -\frac{(pu)^2}{p^2}$
and  Eq.~(A1) may be rewritten as follows:

\bigskip
        $$
        \begin{array}{l}
        \Pi_{\mu\nu}(p) =
        \frac{(\kappa_1 + \kappa_2\lambda)}{(1 +\lambda) (p^2 - (pu)^2)}
        \varepsilon_{\mu\alpha\beta}u^\alpha p^\beta
        \varepsilon_{\nu\xi\phi}u^\xi p^\phi\\
        {}\\
        + \frac{(\kappa_1\lambda + \kappa_2)}{(1 +\lambda) (p^2 - (pu)^2)}
        \frac{(pu)^2}{p^2}
        (p_\mu - u_\mu \frac{p^2}{(pu)} )(p_\nu - u_\nu \frac{p^2}{(pu)} )\\
        {}\\
        + \imath \Re e\alpha_{(1)} \cdot
        \frac{(\kappa_1 - \kappa_2)}{(1+\lambda)(p^2 - (pu)^2)}
        \left(
        \varepsilon_{\mu\alpha\beta}u^\alpha p^\beta
        (p_\nu - u_\nu \frac{p^2}{(pu)} ) -
        (p_\mu - u_\mu \frac{p^2}{(pu)} )
        \varepsilon_{\nu\xi\phi}u^\xi p^\phi                    \right)\\
        {}\\
        - \Im m\alpha_{(1)}  \cdot
        \frac{(\kappa_1 - \kappa_2)}{(1+\lambda)(p^2 - (pu)^2)}
        \left(
        \varepsilon_{\mu\alpha\beta}u^\alpha p^\beta
        (p_\nu - u_\nu \frac{p^2}{(pu)} ) +
        (p_\mu - u_\mu \frac{p^2}{(pu)} )
        \varepsilon_{\nu\xi\phi}u^\xi p^\phi                    \right)
        \end{array}
        \eqno{(A3)}
        $$

\bigskip
$ \lambda = |\alpha_1|^2\frac{p^2}{(pu)^2}$.

\bigskip
The last term in Eq.~($A3$) violates ${\cal PT}$--parity, therefore
the condition $\Im m \alpha_{(1)} = 0$ must hold and $\Pi_{\mu\nu}(p)$
may be
decomposed over the three tensor structures (cf. \cite{salam}). After
exclusion of the last term in Eq.~(A3) the polarization operator may be
presented in a more convenient way:

\bigskip
        $$
        \begin{array}{l}
        \Pi_{\mu\nu}(p) =
        \frac{\kappa_1 + \lambda \kappa_2}{\lambda + 1}
        \left(
        g_{\mu\nu} - \frac{p_\mu p_\nu}{p^2}
        \right)
        + \imath \frac{\lambda^{1/2}}{\lambda +
        1}(\kappa_1 - \kappa_2) \varepsilon_{\mu\nu\alpha}p^\alpha/p \\
        {}\\
         +  \frac{\kappa_1 - \kappa_2}{\lambda + 1}
        \frac{\lambda - 1}{p^2/(pu)^2 - 1}
        \left(
        \frac{p_\mu p_\nu}{p^2} -
        \frac{p_\mu u_\nu + u_\mu p_\nu}{(pu)} +
        \frac{u_\mu u_\nu}{(pu)^2} p^2                          \right)
        \quad .
        \end{array}
        \eqno{(A4)}
        $$

\end{document}